%% file: main.tex
\documentclass[fleqn,usenatbib,useAMS]{mnras}

\usepackage{graphicx}	% Including figure files
\usepackage{float}
\usepackage{amsmath}	% Advanced maths commands
\usepackage{amssymb}	% Extra maths symbols
\usepackage{multicol}        % Multi-column entries in tables
\usepackage{bm}		% Bold maths symbols, including upright Greek
\usepackage{pdflscape}	% Landscape pages

\input{ mnras_packages}

\input{ commands}

\def\cs{\ensuremath{c_{\rm{s}}}}
\def\npeak{\ensuremath{n_{\rm{peak}}}}

\input title

\usepackage[a-1b]{pdfx}
\begin{document}
\label{firstpage}
\pagerange{\pageref{firstpage}--\pageref{lastpage}}
\maketitle
\input abstract

%%%%%%

%optional.
%\input mnras_toc

\input{introduction}

\input{methods}

\input{results}

\input{conclusions}

\input{data}

\input{acknowledgements}

%%%%%%%%%%%%%%%%%%%% REFERENCES %%%%%%%%%%%%%%%%%%

\bibliographystyle{mnras}
\bibliography{bib.bib} % if your bibtex file is called example.bib

%\appendix

%\input{vs_tsing}

% Don't change these lines
\bsp	% typesetting comment
\label{lastpage}
\end{document}

%% file: mnras_packages.tex
\usepackage[T1]{fontenc}
\usepackage{ae,aecompl}
\usepackage{newtxtext,newtxmath}

%% file: commands.tex
\def\EK{\ensuremath{E_{\rm{K}}}}
\def\EG{\ensuremath{E_{\rm{G}}}}
\def\alphav{\ensuremath{{\alpha_v}}}
\def\alphan{\ensuremath{{\alpha_n}}}
\def\AU{\ensuremath{\rm{AU}}}
\def\vrms{\ensuremath{v_{\rm{rms}}}}
\def\pc{\ensuremath{\rm{pc}}}

\usepackage{graphicx, natbib, color, bm, amsmath, epsfig}

\def\sima{\ensuremath{sim1}}
\def\simb{\ensuremath{sim2}}
\def\simc{\ensuremath{sim3}}
\def\tsing{\ensuremath{t_{\rm{sing}}}}
\def\tsung{\ensuremath{t_{\rm{sung}}}}

\newcommand\jcompphys{J.~Comput.~Phys}

\newcommand\revmodphys{Rev.~Mod.~Phys.~}

\newcommand{\half}{\frac{1}{2}}

\newcommand{\avir}{\ensuremath{\alpha_{\rm{vir}}}}
\newcommand{\mach}{\ensuremath{\mathcal{M}}}

\newcommand{\msun}{\ensuremath{M_\odot}}

\newcommand{\dbd}[2]{\frac{ \partial #1 }{ \partial #2} }
\newcommand{\ddbd}[2]{\frac{ d #1 }{ d #2} }

\newcommand{\tff}{\ensuremath{t_{\rm{ff}}}}

\definecolor{orange}{rgb}{1.        ,  0.54,  0}

\definecolor{meta}{rgb}{0.371,0.617,0.625} %cadet blue for
\newcommand{\dc}[1]{}

\def\vvec{\ensuremath{{\bf v}}}

\def\xvec{\ensuremath{{\mathbf x}}}

\definecolor{pink}{rgb}{1.        ,  0.75294118,  0.79607843}
\definecolor{maroon}{rgb}{0.69019608,  0.18823529,  0.37647059}

\definecolor{gray}{rgb}{0.5,0.5,0.5}

\newcommand{\sciform}[1]{\ensuremath{\times {10^{#1}}}}
\newcommand{\percc}{\ensuremath{\rm{cm}^{-3}}}

\newcommand{\kms}{\ensuremath{\rm{km}\ \rm{s}^{-1}}}

\def\cs{\ensuremath{c_{\rm{s}}}}

\def\hw{0.49}

%% file: title.tex
\title[Collapse Histories]{Collapsing molecular clouds with tracer particles:
Part II,  Collapse Histories.}

\author[Collins et al]{David C. Collins,$^{1}$%
\thanks{Contact e-mail: \href{mailto:dccollins@fsu.edu}{dccollins@fsu.edu}}%
Dan K. Le$^1$, Luz L. Jimenez Vela$^1$
\\
$^{1}$Florida State University, Tallahassee, FL 32309}

\date{Last updated 2023 June 15; in original form 2013 September 5}

\pubyear{2023}

%% file: abstract.tex
% Abstract of the paper
\begin{abstract}	
In order to develop a complete theory of star formation, one essentially needs
to know two things:  what collapses, and how long it takes.  This is
the second paper in a series, where we query how long a parcel of gas takes to collapse and the
process it undergoes.  
We embed pseudo-Lagrangian tracer particles in simulations of collapsing
molecular clouds, identify the particles that end in dense knots, and then examine
the collapse history of the gas.
We find a nearly universal behavior of
cruise-then-collapse, wherein a core stays at intermediate densities for a significant fraction of its life before finally collapsing.
We identify time immediately before each core collapses, $t_{\rm{sing}}$, and examine how it
transitions to high density.   We find that the time to collapse is uniformly
distributed between $0.25 t_{\rm{ff}}$ and the end of the simulation at $\sim 1
t_{\rm{ff}}$,
and that the duration of collapse is universally short, $\Delta t \sim 0.1
t_{\rm{ff}}$, where $t_{\rm{ff}}$ is the free-fall time at the mean density.  
We describe the collapse in three stages;
collection, hardening, and singularity.  Collection sweeps low density gas
into moderate density.  Hardening brings kinetic and gravitational energies into
quasi-equipartition.
Singularity is the free-fall collapse, forming an envelope in rough energy balance and central
over density in
$\sim 0.1 \tff$.  
\end{abstract}

% Select between one and six entries from the list of approved keywords.
% Don't make up new ones.
\begin{keywords}
    Stars: formation
\end{keywords}

%% file: introduction.tex
\section{Introduction}

Stars form from clouds of hydrogen that are cold, turbulent, and magnetized.
Temperatures tend to be around 10 K, densities around $100$ \percc, turbulent
velocities a few times the sound speed, and magnetic fields are around a few
microgauss
\citep{Solomon87, Crutcher12, Miville-Deschenes17}.  
\emph{Prestellar cores } are denser and more magnetized \citep{Crutcher12}
objects
that are often associated with embedded infrared sources.  It appears
that a population of these cores are \emph{coherent}, meaning sub- or
trans-sonic \citep{Goodman98, Singh21, Choudhury21}.
The mass distributions of cores \citep{Ladjelate20} and the initial mass
function of stars
\citep{Chabrier03} are similar in shape but offset by a factor of
$\epsilon=3-4$.
This indicates that cores are related to the formation of stars.
One interpretation is that a fraction of each core, $1/\epsilon$, given to each
star \citep{Alves07}, 
though this may be coincidental \citep{Clark07}.

The simplest model for collapsing gas is pressure free collapse in one
dimension.  The density behavior can be well approximated as
\begin{align}
\rho(t) = \rho_0 \left( 1+\left(\frac{t}{\tff}\right)^2\right)^{-a},
\label{eqn.freefall}
\end{align}
where $\tff=(3 \pi/32 G \rho_0)^{1/2}$ and  $a=1.86$ \citep{Carroll06, Girichidis14}.  
This model neglects thermal pressure, magnetic fields, kinetic energy, and
chaotic motions in three dimensions, 
and it produces infinite densities, which is not
physical.
Nevertheless is will be useful in
describing the collapse.  

To improve on this model, \citet{Larson69} and \citet{Penston69} simulated a collapsing sphere
as a radiating hydrodynamic system in one dimension.  They find that the
collapse halts when the \emph{first hydrostatic core} is formed, which happens
when the density and opacity are high enough to trap photons in the core,
causing a dramatic increase in the internal pressure.  These objects are
typically 1-20 AU, 0.05\msun, and have number densities of $n=10^{11}$ cm$^{-3}$
\citep{Young23}.  \citet{Larson69} also showed that the collapse tends to leave
an envelope that has a profile of $\rho\propto r^{-2}$.  

The behavior in one dimension is dependent on its initial conditions.
\citet{Shu77} showed that a sphere will collapse in a self similar manner, and an initial density profile 
$\rho\propto r^{-2}$ will collapse from the inside out.  \citet{Whitworth85}
generalized this self similar behavior, and showed that inside out collapse was a result of the initial
conditions.  These models still neglected the turbulence that has come to
dominate our understanding of star formation.  
Early models had magnetic fields regulating the collapse by slowing
the gravitational contraction \citep[e.g.][]{Shu87, Mouschovias87b}, but these models were largely ruled
out by observations \citep{Crutcher09}.  

The importance of turbulence was first indicated by \citet{Larson81}.  Here it
was found that velocity and size of molecular clouds are correlated
\begin{align}
\vrms &= 1.1 \left(\frac{L}{pc}\right)^\alphav~\kms\label{eqn.larsonv}\\
n &= 3400 \left(\frac{L}{pc}\right)^\alphan~\percc\label{eqn.larsonn},
\end{align}
where $\alphav=0.38$ and $\alphan=-1.1$ were found in that work.  
This has been interpreted to be the
result of turbulence in the interstellar medium
\citep[eg][]{MacLow04,Elmegreen04}. Another interpretation of the observations
puts
gravitational collapse \citep{Ballesteros-Paredes11} as the primary source of
the relationship.  Further, the
consistency of these relations in the Galaxy have been called into question;
\citet{Heyer09} show that the relation depends on the clouds' column densities.
Nevertheless, these relations indicate complex physics at the
cloud level that will dictate the cloud's collapse \citep{Ballesteros-Paredes19,
Xing22,Cahlon24}.

Many modern turbulent fragmentation
models are based on \citet{Padoan95}, wherein some subset of the lognormal
density PDF is turned into stars \citep{Padoan02, Krumholz05, Hennebelle11, Padoan11},
though the assumptions behind these models are different from one another.  More
recent models include the inertial flow model \citep{Pelkonen21}, where large
scale flow drives core formation, and the three phase model of \citet{Offner22},
wherein turbulence gives way to coherent structures.  Our simulation set up and
outcomes are
similar to these, with a few notable differences due to the novel way in which
we examine the collapse.

This is the second paper in a series wherein we inject pseudo-Lagrangian tracer
particles in a simulation of turbulent, collapsing molecular clouds.  This is an extension of \citet[][hereafter Paper I]{Collins23}.  We allow
the cloud to collapse to form several hundred ``stars'', identify the
particles in those ``stars'', and then examine the cells that those particles occupy
during the collapse.  At all times prior to the end of the simulation, we refer
to zones that contain core particles as \emph{preimage} gas.  In Paper I, we examine the preimage gas at $t=0$, the beginning of the simulation.  We find that the density PDF is fully covered by the PDF of
the preimage gas, but suppressed by a factor of roughly
$\left(\rho/\rho_{\rm{max}}\right)^{1/2}$:  all gas
participates in the collapse, dense gas preferentially.  
We examine length
scales of both density and velocity of preimage gas at t=0.  We find that the length scale of the
preimage gas is several times larger than the density auto correlation length, as many density fluctuations feed a single star.
We find that the length scale of the
preimage gas is comparable to the velocity correlation length, indicating one or
a few large scale flows dictate the collapse.
We find that gas is distributed in a
fractal manner, with the Minkowski dimension being distributed between 0.25 and
2, peaking at a dimension of 1.6 (see Section \ref{sec.fractal} for definitions).  We also find a high
degree of overlap of the preimage gas between different cores.  Cores that are close to one another at
the end of the simulation, e.g. binaries, come
from gas that seems to be mixed in the initial cloud.  
Many density fluctuations feed each core, and many cores come
from each collection of density fluctuations.  

In this work, we examine the time history of collapsing gas.  We use the same
simulations described in \citet{Collins23}, but now follow them beyond $t=0$. 
Here we focus on the appearance of self similar collapse from turbulent initial
conditions.  We limit our study to cores that collapse in isolation.
In
future installations of the series  we will discuss magnetic fields, binaries,
and clusters.

The paper is organized as follows. In Section \ref{sec.method} we discuss
the code, simulations, and particle identification.  
Results are presented in Section \ref{sec.results}.  
%We begin by 
%separating cores by their clustering (single, binary, or cluster) in Section
%\ref{sec.modes}. This is necessary to isolate collapse (which we are primarily
%interested in) from tidal effects (which we are also interested in, but is much
%more complex).
In Section \ref{sec.tsing} 
we discuss the collapse time, \tsing, its distribution, and the
unexpected lack 
of correlation between \tsing\ and mean quantities of the preimage gas.  Section
\ref{sec.case_study} follows in detail the collapse of one fiducial object, and we discuss
the three stages of collapse (collection, hardening, and singularity).  We then show the
behavior of the ensemble of isolated objects in 
Section \ref{sec.ensemble}.  
We summarize and
discuss in Section \ref{sec.conclusions}.

%% file: methods.tex
\input{methods_intro}

\input{methods_code}

\input{methods_sims}
\input{methods_pseudocore}

\input{methods_particles}

\input{modes}

%\input{methods_observations}

%% file: methods_intro.tex
\section{Method: Code and Cores}
\label{sec.method}

This work uses the simulations and methods described in detail in \citet[][Paper
I]{Collins23}.  We
will briefly recap the code, simulations, and particle methods here.

%% file: methods_code.tex
\subsection{Code}
\label{sec.code}
We use the open source adaptive mesh refinement code Enzo \citep{Bryan14}
with the constrained transport 
magnetohydrodynamics module \citep{Collins10}.  
This code adaptively adds resolution as the collapse unfolds, when the local
Jeans length $L_{\rm{J}}=c_s^2/(G \rho)^{1/2}$ exceeds
16 zones.
The main solver is a higher order Godunov
method, using the linear method of \citet{Li08} for the differential equations,
the HLLD method of \citet{Mignone07}, and the electric field defined by
\citet{Gardiner05} to maintain the divergence of the magnetic field.  The code
adaptively adds zones using the methods of \citet{Berger89} and
\citet{Balsara01}.  It has been demonstrated to preserve the divergence of the
magnetic field to machine precision \citep{Collins10}, and has been used in a
wide array of applications from star formation \citep{Collins11} to galaxy
clusters \citep{Xu12, Skillman13}.  We additionally include tracer particles that
follow the flow by way of a piecewise linear interpolation of the velocity
field \citep[see][for a description]{Bryan14, Collins23}. Details of particle analysis can be found in Section \ref{sec.particles}.

%% file: methods_sims.tex
\subsection{Simulations}
\label{sec.simulations}

Our simulations begin with periodic turbulent boxes that have been stirred in a manner
described by \citet{MacLow99}.  Here kinetic energy is added to the large scales
of the box, and the nonlinear dynamics carries that energy to small scales.
Driving continues until a steady state is reached, at which time gravity,
refinement,
and tracer particles are added.  We will describe the simulations and then the
physical dimensions of the simulations in the next two sections.

\subsubsection{Simulation Layout}

Ideal magnetohydrodynamics is scale free, and can be characterized by dimensionless parameters.
To characterize our three simulations, we employ the Mach number \mach, virial paramter \avir, and the ratio of thermal
to magnetic energy, $\beta_0$:
\begin{align}
\mach &= \frac{\vrms}{c_s} = 9\\
\avir &= \frac{5 \vrms^2}{3 G \rho_0 L_0} = 1\\
\beta_0 &= \frac{8 \pi c_s^2 \rho_0}{B_0^2} = 0.2, 2, 20.
\end{align}
Here $\vrms$ is the r.m.s. velocity, $c_s$ is the sound speed, $\rho_0$ is the
mean density, $L_0$ is the box size, and $B_0$ is the mean field, taken to be
uniform in the $\hat{x}$ direction.  Simulations continue for slightly less than
one free-fall time of the mean density, $\tff=\left(3 \pi/32 G \rho_0\right)^{1/2}$.

The simulations begin with uniform density and magnetic field and are driven
until steady state is reached.  This was done for three magnetic fields at a
resolution of $1024^3$.  Once
steady state was reached, we downsampled to $128^3$, turned on gravity and
refinement,
and added the tracer particles (one per zone). Refinement 
was triggered whenever the local Jeans length became smaller than 16 zones, $L_J=c_s/(G \rho)^{1/2}<16 \Delta
x$. This continued for 4 levels for a maximum effective resolution of $2048^3$.  
%\red{While this resolution is lower than what we have performed in the past,
%Troles found convergence, so at least its not too bad.}

\subsubsection{Physical Parameters}

While we can formally select the physical units to be what we like, subject to
the above dimensionless ratios, it is useful to select a fiducial scale.  To do
so we assume that our simulations are consistent with Larson's relations
\citep{Larson81}, see Equations \ref{eqn.larsonv} and \ref{eqn.larsonn}.
We must also adopt a sound speed. We
assume $T=12$K, which gives a sound speed  of $\cs=0.2~\kms$.

Thus the physical scale for velocity, length, density, mass, global free fall
time, and smallest cell size are found to be: 
\begin{align}
\cs &= 0.2\ \kms\\
\vrms &= 1.8 \left(\frac{c}{\cs}\right)\ \kms\\
L_0 &= 3.6\left(\frac{c}{\cs}\right)^{1/\alphav}\ \pc\\
n_0 &= 817 \left(\frac{c}{\cs}\right)^{\alphan/\alphav}\ \percc\\
M &= 2267 \left(\frac{c}{\cs}\right)^{(\alphan+3)/\alphav}\ \msun\\
\tff &=  1.2 \left(\frac{c}{\cs}\right)^{\alphan/2\alphav}\ \rm{Myr}\\
\Delta x &= 368\left(\frac{c}{\cs}\right)^{1/\alphav}\ \AU.
\end{align}

%% file: methods_pseudocore.tex
\subsection{Pseudo-cores and Sink Particles}
\label{sec.pseudo}

The first hydrostatic core forms when the opacity of the gas grows large enough
to stiffen the equation of state from isothermal \citep{Larson69}.  These
objects are only $\sim 20$ AU across, and as such not resolvable in our
simulations.  An approximation must be made.  Most often
sink particles are employed, which remove some mass from the grid and replace it with
a particle that has zero volume \citep{Federrath10b, Teyssier19}.  This is the
industry standard for dealing with the large densities.

We have chosen to not use sink particles for two reasons.  First, the creation
of the sink disrupts the very process we are trying to measure, since the
algorithm will remove mass, kinetic, and thermal energy from the collapsing
object.  Second, there are presently no sink algorithms that properly treat
magnetic fields, each opting to ignore \citep{Teyssier19}.  This violates Alfven's theorem that we work so hard to enforce, and the consequences on the collapsing dynamics have yet to be studied.

Instead we allow the grid pressure to supply the additional support that would
be provided by thermal pressure.  This allows us to examine the onset of self
similar collapse and examine the precipitating prestellar core without the
interference of other algorithms.  This is philosophically similar to the idea of implicit
large eddy simulations \citep{Adams09}.  In an such a simulation, the grid
viscosity is used in place of an explicit sub grid model of the unresolved
scales.  The basic idea is that the nature of the turbulent cascade does not
depend on the nature of the dissipation, only its existence.  
In a similar manner, the collapse of a 
collapsing molecular cloud will not depend on the nature of the approximation we
make at the end, within reason.  We must then use care in dealing with the numerical artifacts
that are leftover, regardless of their nature.

At the end of the collapse, 
we form small (a few zones across, $\sim 1000$ AU)
high density ($n\sim 10^{5} n_0 \sim 10^{8} \rm{cm}^{-3}$) condensations that
we refer to as \emph{pseudo-cores}.  
They're several hundred times larger than the first core they aim to replicate
and thus not numerically resolved. Therefore we refrain from quantitative analysis of the
population of objects, and we cease detailed analysis of each after they have formed.  For this reason, among others, we restrict our analysis to isolated cores that do not interact, as it is clear when analysis should be stopped for each core.

To identify pseudo-cores, we use a dendrogram type technique \citep{Rosolowsky08, Turk11} to
find the location of dense peaks.  We keep only the peaks above $10^4 n_0$, as most dense peaks have
$\npeak\sim 10^{5-7}n_0$, while turbulence only makes peaks with
$n<10^3n_0$.  There are few fluctuations with $n\sim 10^4n_0$, these
are either turbulent fluctuations or too early in their collapse to learn anything from.  For
each peak, we take the particles closest to the densest zone.  For practicality,
we take all the zones such that $n>\npeak^{3/4}$ around each peak, but
the definition of which zones to take does not matter substantially as the
particles cluster in the densest few zones.  We refer to the pseudo core and the
particles within collectively as a \emph{core}, as we are interested in the prestellar core around each pseudo-core, and in this analysis each one is a unique path in space.  We refer to zones that are flagged as containing core particles at
earlier times as \emph{preimage} zones.

In our analysis we focus on two things; the process that delivers mass to the pseudo-core, and the \emph{envelope} around the core.
The $\sim0.1$pc envelope around the pseudo-core would potentially be considered a prestellar core, but we do not tie ourselves to these observationally defined objects.

It is worth emphasizing that a pseudo-core is not a resolved object, we
introduce them here only to isolate the unresolved endpoint of the collapse from the
collapse itself and the remaining envelope in our presentation of the results.

%% file: methods_particles.tex
\subsection{Particles}
\label{sec.particles}

Tracer particles are evolved using a basic kick-drift approach, where the
particle velocity $\vvec_{\rm{p}}$ is interpolated from the grid velocity in a linear
manner, and the position is then updated as $\xvec^{n+1} = \xvec^{n} + \Delta t
~\vvec_{\rm{p}}$.   Particles are initially deposited uniformly throughout the box, one
per zone.  

Particles
have a density (defined by the inter particle spacing) and velocity
(interpolated from the grid) that are not of interest.
We only use the particles to determine which gas zones to analyze.   One of two
methods is used:  either the zones containing the particles (i.e. preimage gas),
or a sphere defined by the centroid and extent of the particles is used.
The sphere has a minimum radius of one root grid zone, as the particles eventually
occupy only a few fine-grid zones.

Particles, unlike gas, do not feel each others' presence or pressure of any kind.
One result of this is over clustering
of the particles relative to the gas, which results in a much wider distribution
of particle densities \citep{Genel13}.
The reason for this is that two particles, once within the same
zone in a converging flow,  will be driven to zero separation in a short time
\citep[see][]{Collins23}.  Particle velocity is linearly interpolated from the
grid velocity, and as such it is unlikely to separate the particles as the
nonlinear turbulent field would be were the subgrid velocity resolvable.   
This results in a loss of the number of
degrees of freedom the particle flock has as the collapse proceeds.  For our
simulations, this is a useful feature rather than a problem.  It means that the
particles do not inject any noise of their own, so we can reliably trace them
back to their origin.  We take care to not perform any measurements that rely on
the information in the particles after they collapse and all occupy the same few
zones.  

The net effect of the over clustering is in fact a useful feature of the
particles for this study.  Once the gas has gotten dense and many particles
occupy a single zone, their individual contributions no longer have useful
information. By sticking together they effectively remove themselves from
consideration as individual pieces of information, eliminating confusion they
may cause.

On remedy for the over clustering that has been suggested is to add some effective
kinetic temperature to the particles, which would allow the particles to entrain
with the turbulence more effectively \citep{Genel13}.   For this study, adding
noise to the particle trajectories would destroy the causal connection between the
particle at the beginning and the end of the simulation, and is not desirable.

%% file: modes.tex
\subsection{Mode Separation}
\label{sec.modes}

We begin our analysis by classifying  the collapse mode of each core;
\emph{single},
\emph{binary}, or \emph{cluster}.  
This allows us to avoid the effects of unresolved objects and tidal interactions.

For each core, we measure the distance, $\delta$, 
between its centroid and the centroid of all other cores for all time steps.  The number of cores
within $\delta=0.05 L_0$, $N_\delta$, indicates the mode of collapse; single
($N_\delta=1$), binary ($N_\delta = 2$) or cluster ($N_\delta>2$).  The value of $\delta$ was determined by examining the path of the centroid for each core and its neighbors.  The results presented here are insensitive to the exact value of $\delta$.
This simple grading scheme was
verified by visual inspection of movies of the collapsing cores.
While more sophisticated methods for identifying neighbors can be concocted,
this simple scheme is sufficient for our purposes here.
Table \ref{table1} shows the number of cores in each mode for each simulation, though these demographics are as approximate as our separation scheme.

%Figure \ref{fig.hair} shows the collapse path in space for several
%representative cores; two single, one binary, and one cluster.  This figure
%shows projected $x$ and $y$ position relative to the total box, $L_0$.  As the domain
%is located between 0 and $L_0$, negative values represent periodic wrapping.
%A track for each particle begins at the dots and end at the
%red point indicated by the number.  The first three panels show every particle for
%each core, the bottom right panel shows only the centroids of 
%each core in the cluster.  The top row shows core 214 and 74, which both form alone.
%Core 214 forms from a disjoint set, which is spatially a fractal, see
%\citet{Collins23} for a more detailed discussion of fractal preimages.  Core 74 appears to form from a more
%continuous region of space, but this is merely a projection effect, it too does
%not fill the space and has a fractal dimension of 1.6.  In the bottom left,
%cores 112 and 113 begin with their particles mixed, and the gas can be seen to
%"unmix" as the collapse proceeds.  The bottom right panel shows a cluster of 5
%cores, (367, 368, 369, 370, and 371) which all form  and evolve near each other.
%Initially the gas forms in one basin, with the preimage gas mixed together.
%For clarity, we plot only the centroids of the paths.
%The orbital dynamics of the cores interacting with one another is clear from the
%braided paths.  

%% file: results.tex
\section{Results}
\input{table1}
\label{sec.results}

\input{results_intro}

\input{fig_density_only}

\input{fig_tsing_dist}

\input{tsing}
\input{table2}

\input{fig_anatomy}

\input{case_study}

\input{fig_violent}
\input{violent_relaxation}

\input{ensemble}
\input{fig_velhair}
\input{fig_r1000}
\input{fig_freefall}

\input{fig_fractal_time}
\input{fractal_time}

\input{fig_radial_prof}

\input{radials}

%% file: table1.tex
\begin{table} \begin{center} \input{table1_caption} \label{table1}                                                                                       
\begin{tabular}{ r l l l }
    \hline
  & \sima & \simb & \simc \\
  \hline
    $\beta_0$ & 0.2 & 2 & 20 \\
\hline

Single & 47 & 17 &  50 \\

Binary & 44 & 26 &  25 \\

Cluster & 22 & 69 &  61 \\

\hline
\end{tabular}                                                                                       
\end{center}                                                                                       
\end{table}       

%% file: table1_caption.tex
\caption{Mode of star formation by simulation.}

%% file: results_intro.tex
Now we describe the behavior of the collapse.  In Section
\ref{sec.tsing}, we define and measure \tsing\ and \tsung, which indicate the beginning and end of the free-fall collapse. 
We also measure the distribution of collapse
times, showing a universally short collapse, faster than free fall of the mean
cloud.
We focus our attention on  a careful examination
of one representative core in Section \ref{sec.case_study}.
We then show this same behavior for all single cores in
Section \ref{sec.ensemble}.

%% file: fig_density_only.tex
\begin{figure}[h] \begin{center}
	\input{fig_density_only_img}
\caption[ ]{Density vs time for two representative cores.  Core
214 collapses much more slowly than 74. In each panel, the
first grey line represents \tsing while the second grey line represents \tsung.}
\label{fig.rho_t} \end{center} \end{figure}

%% file: fig_density_only_img.tex
\includegraphics[width=\hw\textwidth]{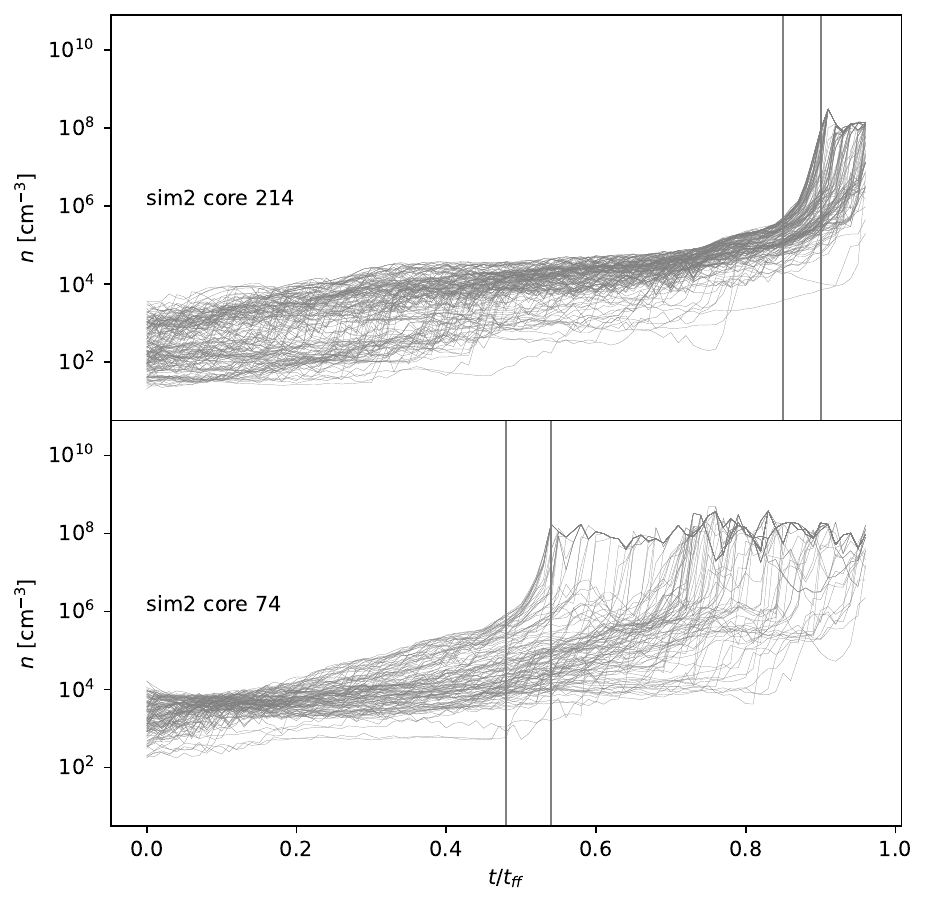}

%% file: fig_tsing_dist.tex
\begin{figure} \begin{center}
	\includegraphics[width=\hw\textwidth]{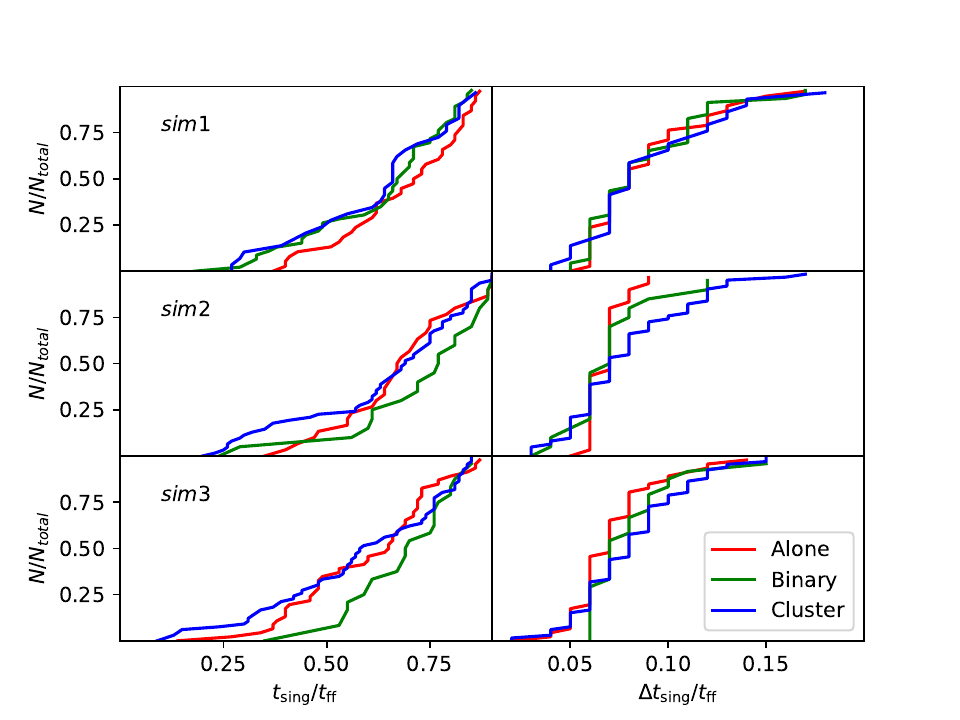}
	\caption[ ]{(\emph{Left})The cumulative distribution of the singularity
	time, \tsing. For each mode of collapse, the singularity time is nearly
	uniformly distributed.  (\emph{Right}) The duration of the singularity,
	which is distributed between $0.05 \tff$ and $0.15 \tff$.}
\label{fig.tsing} \end{center} \end{figure}

%% file: tsing.tex
\subsection{Singularity Time: How Fast}
\label{sec.tsing}

\subsubsection{The singularity time, \tsing}

Figure \ref{fig.rho_t} shows density vs. time for two cores.
Core 214 is quite
slow, not collapsing until near the end of the simulation, while  core 74 begins to
get dense much earlier, with a few particles reaching high density early, and
the remaining particles accreting over the remainder of the simulation.   
It is useful to identify the moment immediately before the collapse, which we
denote \tsing, and the moment when the collapse ends, which we denote \tsung.
These are found by examining when the time derivative of the maximum of density
over the preimage gas; \tsing\ is defined as the time when the derivative gets above a threshold value,
and \tsung\ is defined as the time when the density derivative is zero again.
That is,
\begin{align}
	\dbd{n_{\rm{max}}}{t}\biggr\rvert_{\tsing} &=10^5
	\frac{n_{\rm{code}}}{t_{\rm{code}}} =
	7\sciform{7}\frac{\rm{cm}^{-3}}{\rm{Myr}}\\
	\dbd{n_{\rm{max}}}{t}\biggr\rvert_{\tsung}&=0,
\end{align}
where the threshold value is found to be $10^5$ in code units, or $7\sciform{7}
\rm{cm}^{-3}\rm{Myr}^{-1}$ using the scaling defined earlier.  
These are indicated by vertical lines in Figure \ref{fig.rho_t}.

The cumulative distribution of \tsing\ for all cores can be seen in the left column of Figure \ref{fig.tsing}.
Each row shows a different simulation.  Red lines show single cores, green shows
binaries, and blue shows clusters.  For each population, the cumulative
distribution indicates that the distribution of \tsing\ is relatively uniform
between $0.25\tff$ and $1 \tff$ when the simulation ends.  Some fluctuations
exist, but no consistent trends are noticeable.  

The right column of Figure \ref{fig.tsing} shows the distribution of the
duration of the initial singularity, $\Delta \tsing=\tsung-\tsing$.  Note
that the horizontal axes between the left and right columns are not the same.  The collapse shows a nearly universal
short duration.

We can explain the typical behavior of $\Delta \tsing$ by noting the almost
free-fall behavior at the end of the collapse.  The typical density of the
preimage gas at \tsing\ is 500$n_0$, and the grid pressure sets in at roughly
$10^5n_0$.  Free fall is self similar, so the
timescale to go between two densities is proportional to the square root of the
densities.   This  gives a collapse time of $\sim0.1 \tff$ to go from the
density at \tsing\ tolthat at \tsung.  This does not, though, imply a critical
threshold for collapse; by the time the gas reaches the density it has at
\tsing\ the collapse has already been triggered.

\subsubsection{What sets \tsing?}
\label{sec.what}

It would be extremely useful to understand what made some cores collapse faster
than others.  Unfortunately, after examining many possible correlations, we do not find any that show strong correlations with the collapse time.  For each simulation, we examine Pearson $R$ for \tsing\ vs. several properties of the cores.
Pearson $R$ measures the
degree of linear correlation between two quantities, with -1 being perfectly
anti-correlated.  

 $R$ is defined as
\begin{align}
    R = \frac{\langle Q \tsing \rangle_c}{\sqrt{\langle Q^2\rangle_c\langle
    \tsing^2\rangle_c}}.
\end{align}
Here the average $\langle \rangle_c$ is taken over all cores, and $Q$ is the average or total of the quantity over the preimage gas for each core.   For the extrinsic quantities, $Q$ represents the total over preimage, which includes the number of particles, $N_{\rm{particles}}$, and the volume of the convex hull bounding the preimage, $V_{\rm{hull}}$.  
For other quantities, $Q$ is the mass-weighted average of the quantity over the
preimage gas for core $c$.   

We examine the correlation coefficient for the following quantities: the number
of particles $N_{\rm{particles}}$; the volume of the convex hull bounding the
particles $V_{\rm{hull}}$; the mean density, $\overline{ \rho}$; the mean
gravitational binding energy, $\overline{ \EG}$; the mean kinetic
energy, $\overline{ E_K }$;  the r.m.s. velocity,
$\overline{v_{\rm{rms}}}$; the mean radial velocity,
$\overline{ v_{\rm{R}}}$; the mean tangential
velocity $\overline{ v_{\rm{T}} }$; and the free-fall-time at the mean density of the
core, $\overline{\tff}=(3
\pi /32 G \overline{ \rho })^{1/2}$.   

Table \ref{table2} shows $R$ for these quantities for each of the three
simulations.  Only the single cores are included in this table.  
There is only one correlation that is as high (in magnitude) as 0.5, $\overline{\rho}$ for \simc,
but only a mild anti-correlation is seen in the other two simulations, with \sima\ and
\simb\ having values of -0.34 and -0.33, respectively.  The number of particles
has a mild trend for anti-correlation (-0.41, -0.37, and -0.24) for the three
simulations, respectively.  The local free fall time shows a mild positive
correlation,
with coefficients of (0.37, 0.38, and 0.48), respectively.  The velocity and
energies are all weakly correlated, with no consistent trend in the simulations.

While the trends are in the direction one would expect (e.g. \tsing\ is
positively correlated with $\overline{\tff}$, and the velocity dispersion is
negatively correlated) the scatter is too large to be predictive.
This is not to say that $\tsing$ is not predictable, but the obvious things to
check are individually not able to predict \tsing.

%% file: table2.tex
\begin{table} \begin{center} \input{table2_caption} \label{table2}                                                                                       
\begin{tabular}{ r l l l  }
    \hline
Q & \sima & \simb & \simc \\
\hline

$N_{\rm{particles}}$ & -0.41 & -0.37 & -0.24   \\

$V_{\rm{hull}}$ & 0.08 & 0.02 & -0.21   \\

$\overline{ \rho }$ & -0.34 & -0.33 & -0.50   \\

$\overline{t_{\rm{ff}}}$ & 0.37 & 0.38 & 0.48   \\

$\overline{ \EG}$ & -0.001 & -0.28 & -0.30   \\

$\overline{ \EK }$ & 0.04 & -0.44 & -0.10   \\

$\overline{v_{\rm{rms}}}$ & 0.15 & -0.15 & 0.08   \\

$\overline{ v_{\rm{R}} }$ & 0.13 & -0.15 & 0.08   \\

$\overline{v_{\rm{T}}}$ & 0.11 & -0.17 & -0.001   \\

\hline
\end{tabular}                                                                                       
\end{center}                                                                                       
\end{table}

%% file: table2_caption.tex
\caption{Pearson R between properties, Q, for each single core and their singularity
time, $\tsing$.  No substantial correlations are seen.}

%% file: fig_anatomy.tex
\begin{figure} \begin{center}
	\includegraphics[width=\hw\textwidth]{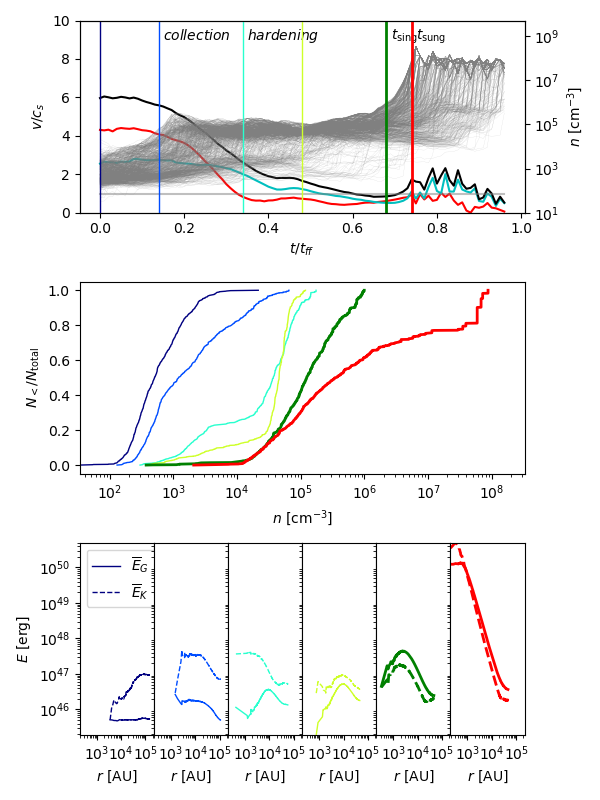}
\caption[ ]{The collapse of one core and its phases.  \emph{(Top row)} Density (grey
lines, right axis) and velocity (radial in red, tangential in teal, and total in black, left axis).
Vertical lines denote the times of analysis for the second and third rows.
  These are the beginning (navy) the middle of the hardening phase
(light blue), the end of the hardening phase (light green), the beginning of
singularity (dark green) and the ``end'' of singularity (red). \emph{(Middle row)} Cumulative density distribution at the 5 indicated
snapshots. \emph{(Bottom row)} Radial profiles of of $\overline{E}_G$ and
$\overline{E}_K$ for each of the
5 snapshots. }
\label{fig.anatomy} \end{center} \end{figure}

%% file: case_study.tex
\subsection{Collapse Paradigm}
\label{sec.case_study}

Here we examine one core in detail.  
To further understand the processes at work, we would like to examine $n_i(t)$ for every
particle, $i$, in every core, as well as velocity, kinetic, and gravitational
energies.  This is several thousand plots.  We summarize these plots by way of a
case study (this section), followed by an ensemble of cores (next section).  

\subsubsection{Three stages}

We find that each core goes through three \emph{stages} of collapse to one degree or another.
We refer to these as \emph{collection}, \emph{hardening}, and
\emph{singularity}.
The collection stage sweeps low
density gas together by converging supersonic flows; kinetic energy is typically
larger than gravitational, but not universally.  This stage does not appear to
happen for every core.  During the hardening stage, the
density becomes moderately high ($\sim 10 n_0$), the velocity decays to
trans- to slightly super-sonic, and gravitational and kinetic energies
equilibrate.  During the singularity, density rises extremely fast and the
majority of the mass is delivered to the core in $\sim 0.1 \tff$; gravitational
energy dominates.  By the end of the singularity, the envelope has a density
profile proportional to $r^{-2}$, but
not generally spherical, and the ratio of kinetic to gravitational energy
follows a remarkably consistent pattern dropping from $\EK/\EG\sim4$ at the
inner radius to $\EK/\EG=0.5-2$ at outer radii.

\subsubsection{Case Study}

In Figure \ref{fig.anatomy} we show the collapse history of one representative
core (core 114
from $sim2$) outlining the three stages of collapse.  
This plot summarizes how density, velocity, gravitational energy, and
kinetic energy change in concert with one another.  The top row shows the
preimage density, one line
for each particle
(grey lines, right axis) and the average velocity (r.m.s in black, radial velocity
in red, and tangential velocity in teal, left axis).  Also indicated are
 six snapshots in time (vertical bars).  The second row shows the
cumulative density distribution for preimage density at each of those snapshots.  The third row shows radially averaged
kinetic and gravitational energy for a sphere surrounding the particles at each snapshot. 
The first two rows utilize only the preimage gas, while
the third row measures $\EG$ and $\EK$ on spheres centered on the centroid of
the particles. 
We will elaborate on each of
these in turn.  

Six snapshots in time, indicated by the vertical bars in the top row of Figure
\ref{fig.anatomy}, highlight the stages of the collapse;
   the initial conditions (navy), the collection stage 
(light blue), the  hardening stage (teal and light green), the beginning of
singularity (\tsing, dark green), and the end of singularity (\tsung, red). 
We do not
presently have a quantitative boundary between \emph{collection} and
\emph{hardening}, as
no consistent trends are obvious (see Section \ref{sec.ensemble}).  

The top row in Figure \ref{fig.anatomy} also shows the density and velocity
behavior of this core.  Density begins low, sampling the turbulent density
field, as discussed in Paper I.
The total velocity is initially quite large, $v_{rms}=6
c_s$ in this core (not universal).  During the collection stage (blue lines), the large scale flow collects the material into a
smaller, higher density patch.  During the hardening stage (teal, light
green lines), the density
distribution becomes more narrow and the velocity decays.  For this core, the
radial velocity becomes transsonic by $t=0.3\tff$, at
which time the tangential velocity increases over the
radial.  The r.m.s velocity is subsonic at the start of singularity, but this is
also not a universal feature among all cores.  After the end of the singularity,
the core accrets the remaining particles.  
As we discuss in Section \ref{sec.violent}, 
this stage does not accrete much mass, the vast majority of the mass is
assembled during the singularity.

The second row in Figure \ref{fig.anatomy} shows the cumulative preimage density
distribution at each of the six snapshots.  At the first time (navy line), the density is roughly
lognormal, as it samples the turbulent initial conditions.  This core begins in
a large converging flow
which sweeps most (80\%) of the
particles to moderate density
($n=10-100 n_0$), while some remains lower ($n=10n_0$).  During the
\emph{hardening} stage, the moderate-density gas gets moderately more dense, and
the low density gas is compressed, but the peak density does not increase
dramatically.  During the singularity (between the dark green and red lines),
the peak density increases rapidly.  Roughly 20\% of the particles are above a density
of $10^7 \percc$ at the end of the collapse. 

The third row in Figure \ref{fig.anatomy} shows the gravitational and kinetic
energies of the gas on spheres surrounding the particles.  The dotted line shows
kinetic energy, averaged over the sphere of volume $V(r)$, and the solid line
shows the magnitude of gravitational
energy: 
\begin{align}
\overline{E}_{\rm{K}}(r) &= \frac{1}{V(r)}\int_0^r \half \rho (v-\bar{v})^2 d^3x\nonumber\\
\overline{E}_{\rm{G}}(r) &= \frac{-1}{V(r)} \int_0^r \frac{\left(\nabla \phi\right)^2}{8 \pi G} d^3x\label{eqn.EGEK}\\
\end{align}
where $V(r)$ is the volume of a sphere with radius $r$.
Initially the gravitational energy is flat and kinetic is several times
larger.  By the end of the hardening stage (light green line) the profiles of
\EG\ and \EK\ have been driven toward each other.  Singularity drives an
enormous increase in both kinetic and gravitational energies, and we're left
with a relaxed envelope with a rapidly rotating pseudo-core in the center.

%% file: fig_violent.tex
\begin{figure} \begin{center}
\includegraphics[width=0.5\textwidth]{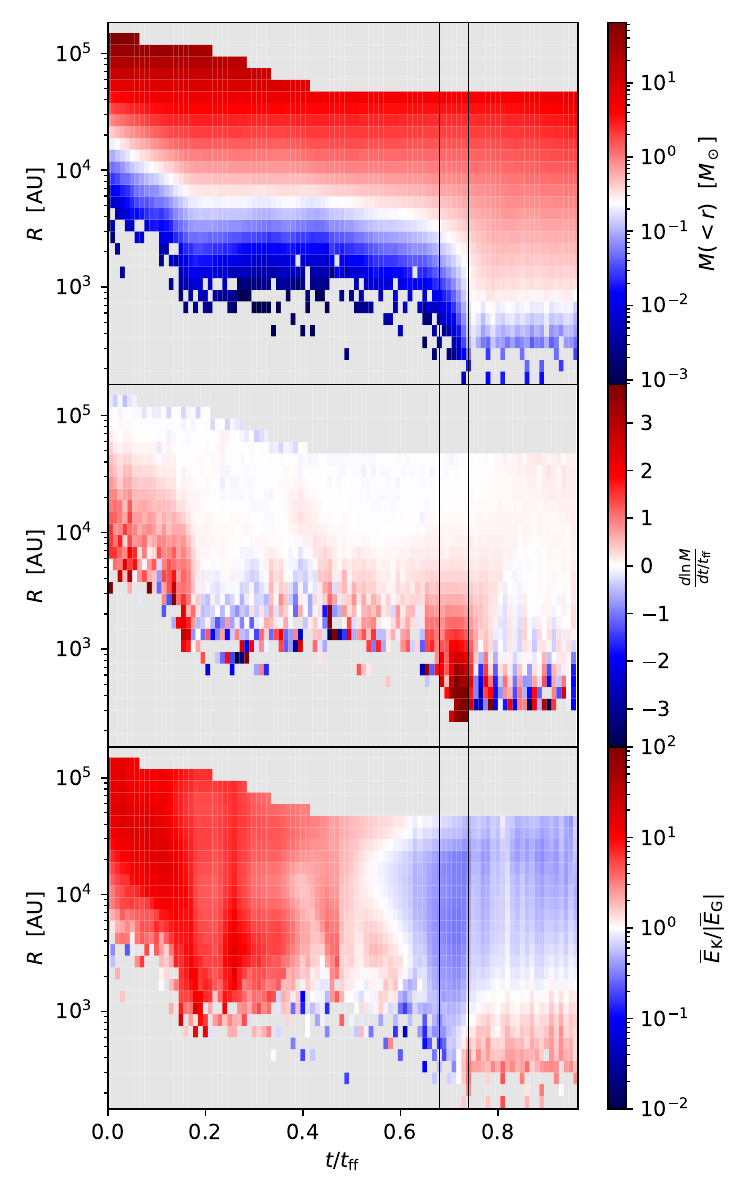}
\caption[ ]{Extremely rapid assembly of mass.  All three plots show azimuthally
averaged quantities on a sphere centered on the particles, but taking all zones.
The left vertical axis is radius in AU, the horizontal axis is time in units of the free
fall time.
Vertical black lines show \tsing\ and \tsung.  (\emph{Top}) Interior mass,
$M_<$.
The mass is assembled to the center extremely rapidly between \tsing\ and
\tsung, with relatively little accretion after \tsung.  (\emph{Center}) The
relative mass accretion rate.  Very little accretion happens after \tsing.
(\emph{Bottom}) The ratio of kinetic and gravitational energies.  Immediately
after \tsung, the energies show that the core forms a relaxed envelope. During the singularity, gravitational
energy dominates. }
\label{fig.violent} \end{center} \end{figure}

%% file: violent_relaxation.tex
\subsubsection{Mass Delivery}
\label{sec.violent}

\def\rsph{\ensuremath{R_{\rm{sphere}}}}
We now examine the nature of the gas mass during the collapse.  Figure
\ref{fig.anatomy} only shows density of the preimage gas, which covers a sparse subset
of the surrounding material.  It is valuable to also make a complete census of the gas around the
collapsing object.  

Figure \ref{fig.violent} shows mass vs. time and radius.
At each time, a sphere
of gas is extracted that is centered on the centroid of the particles and has a
radius equal to the most distant particle.  The particles all collapse
to a few zones at the end, so the analysis sphere has a minimum radius of one
root-grid-zone. The vertical lines show \tsing\ and \tsung.  

The top panel shows interior mass vs time,
\begin{align}
M(<r)&=\int_0^r \rho d^3x\\
&:= M_<\nonumber
\end{align}
The edge of a clump is poorly defined, so to avoid murky clump edge definitions we adopt a fiducial radius of $1000$AU and fiducial mass,  $M_{1000}$, the mass within that radius at the final snapshot.
The colorbar is arranged so that the white line follows a mass $M_{1000}$. It should
be noted that this is not, strictly speaking, a Lagrangian surface, since there
is substantial flux both in and out of this surface.  
It does show that the mass
is assembled at the center of the core rapidly during the singularity, and then
remains relatively constant.

The middle panel shows the normalized relative accretion rate:
\begin{align}
    q &=  \ddbd{\ln M_<}{t/\tff}\\
    &= \frac{\ddbd{M}{t}}{M/\tff}.
%&= \tff \frac{\int \rho v_r d^2x}{M}.
\end{align}
This is computed 
by way of finite differences of the binned $M_<$ in the top plot.
We see a large positive accretion rate during the collection phase
($t<0.2\tff$), followed by a high degree of volatility, fluctuating around zero.
Beginning at \tsing\ and
ending at \tsung, $q$ grows to a factor of several as most of a solar mass is
delivered to the center in 0.1 \tff.  Then the high frequency rotational motions
of the pseudo-core make the mass fluctuate in its location, but not gain much overall
mass.  The lack of growth in mass is due to the fact that the envelope is
virialized,  with large kinetic energy.

The third panel of Figure \ref{fig.violent} shows the ratio of kinetic to
gravitational energy,
$\EK(r)/|\EG(r)|$, both defined in Equation \ref{eqn.EGEK}.  
The colorbar is tuned so that $\EK(r)/|\EG(r)|=1$ is white; gravitationally
dominated gas is in blue, and kinetic energy dominated gas is in red.  
It is seen that at the beginning of the simulation, kinetic energy
dominates the bulk of the flow.  As time progresses during the \emph{hardening}
phase, the ratio grows to roughly unity.  During the singularity, gravity
dominates the entire system, and immediately after \tsung\ we find a pattern
slightly dominated by kinetic energy ($\EK/|\EG|=2$),  and transitions to
$\EK/|\EG|=0.5$ in the outer regions.

It should be noted that these dynamics are visible to us because we did not use
sink particles.  Sink particles \citep{Teyssier19} are a useful tool to follow
self-gravitating simulations in time, but they are a subgrid model selected by
the simulator, and the accretion rate they give is, while generally reasonably
constructed, dramatically influenced by the parameters of the sink.  Here we allow the dynamics to unfold self-consistently, at the
cost of shortened simulation time.  This allows us to measure how singular
collapse from turbulent initial conditions unfolds.

%% file: ensemble.tex
\subsection{Ensemble properties}
\label{sec.ensemble}

The previous section examined one core in detail. Now we examine a larger sample of cores to see
how many of these traits are universal, and how many have some variation.
We restrict our examination here to \emph{single} cores, from the first
simulation \sima.  
Binary cores, and cores from the other simulations, 
follow
similar trends.  A fraction of cluster cores do as well 
but many
cluster cores have much more complex velocity signatures and are likely not
actually relaxed at the end of the simulation.  We revisit binaries and clusters
in a future work.

\subsubsection{Mean density and velocity of preimage gas}

Figure \ref{fig.velhair} shows mean of preimage gas quantities for every core in
the sample.  Each core has been scaled in time to its own \tsing.  The top row
shows mean density, $\overline{n}$, which shows a universal behavior of slow increase followed by a
transition to a free-fall solution (cruise-then-collapse).  The second row shows the r.m.s. velocity of
the preimage gas, $\overline{v_{\rm{rms}}}$.  The flow begins highly supersonic in total.  Velocity decays
by \tsing\ to a Mach number of a few.  The third row shows the mean radial
velocity, $\overline{|v_{\rm{R}}|}$. 
Interestingly, there is not a universal trend of large converging flow. In fact,
some cores have very small radial velocities initially.  To the extent that the
collection stage is real, this can be seen in this plot as large radial
velocities that decay to transsonic to slightly supersonic.  Finally, note that
this is the absolute value of the radial velocity, but all mean velocities are
negative.  By \tsing, the
radial velocity is clustered around transsonic.  Some are subsonic, as expected
by others, while some are slightly supersonic.  The fourth row shows the tangential
velocity, $\overline{|v_{\rm{T}}|}$,  which is universally supersonic initially, and decays to transsonic
before \tsing.

\subsubsection{Mass Delivery}

The fast mass delivery for each core can be seen in Figure \ref{fig.r1000}.
Here we define $R_{1000}$ as the radius that contains a mass of $M_{1000}$,
which is defined as the mass within 1000 AU at the end of the simulation (see Section \ref{sec.violent}.)     This can be seen as the white line in the top panel of Figure \ref{fig.violent} for one
core.  In Figure \ref{fig.r1000}, we show $R_{1000}$ for each core vs. time,
scaled to the cores own \tsing.  Most cores deliver the bulk of the mass of the
final object beginning at \tsing and taking about 0.1 \tff\ to get there.  A few
behave slightly differently, as these are ultimately chaotic dynamics.  Note that this plot measures all gas, not just
preimage gas.

\subsubsection{Scaled Free Fall}

We find that density is reasonably described by \emph{scaled free fall}.  Given an
isolated pressure free sphere of gas collapsing under its own gravity, it's radius can be computed as a function
of time.  This yields a transcendental equation for radius that can be
approximated by 
\begin{align}
    n(t) &= n_0 \left( 1-\left[\frac{t}{\tff}\right]^2\right)^{a/3},\label{eqn.ff2}
\end{align}
where $a=1.8614$ \citep{Girichidis14}.  We find that the density of a collapsing
object is reasonably
approximated by \emph{scaled free fall}, 
\begin{align}
n_{\rm{s}}(t) &= n_0 \left( 1-\left[\frac{t}{\tsung}\right]^2\right)^{-a}.
\label{eq.freefall}
\end{align}
This is faster than free fall, as $\tsung<\tff$ for all cores in our study.  

Figure \ref{fig.freefall} shows scaled free fall and the renormalized density
for each core.  The top panel shows the maximum of the density for each core,
normalized to its value at the start, and the time is stretched to the end of
the singularity, \tsung:
\begin{align}
    \frac{n_{\rm{max}}(t/\tsung)}{n_{\rm{max}}(t=0)}.
\end{align}
This is plotted for each core, along with Equation \ref{eqn.ff2}.  Scaled
free fall forms a lower bound for the suite of curves; clearly free fall is not
the only physics at work, but it serves as the stage on which the dynamics
unfold.  

The lower panel of Figure \ref{fig.freefall} shows a similar plot, but with the
maximum over density replaced with the mean of the preimage density.  The mean
densities do not decrease past $t=0$ the way the maximum density does, and the
curves are all somewhat denser that what free fall at this rate would collapse.
Again, free fall collapse is not the only physics at work, but is a useful
framework.

If a useful prediction of \tsung\ could be found, this could be used in a 
predictive theory of star formation.

%% file: fig_velhair.tex
\begin{figure} \begin{center}
\includegraphics[width=\hw\textwidth]{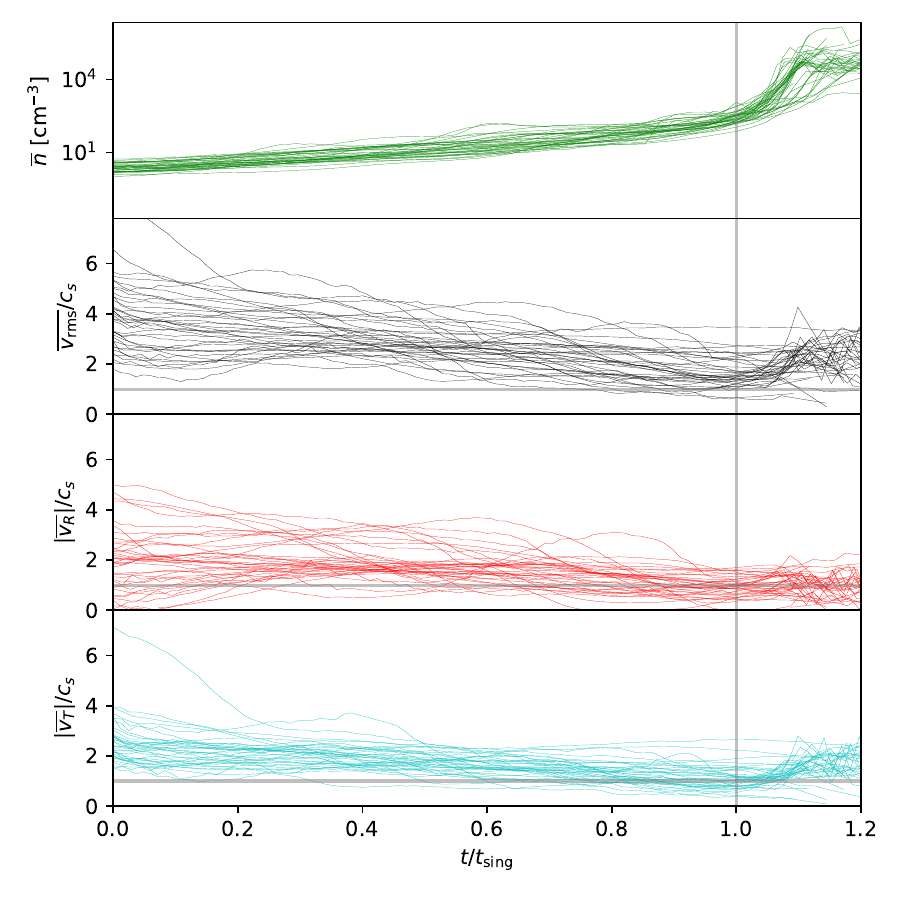}
\caption[ ]{Tracks for all of the \emph{single} cores from \sima, each one
scaled to its respective \tsing.  Each line represents a core.  \emph{(Top Row)} Mean density of preimage gas.  A
consistent profile can be seen of slow increase followed by free-fall.
\emph{(Second Row)} r.m.s velocity of preimage  gas.  The total velocity
remains quite large.  \emph{(Third Row)} Radial velocity of the preimage.  Radial
velocity has no preferred initial configuration.  Some begin with large
convergence, some do not.  By \tsing, the radial velocity is clustered around
transsonic.  \emph{(Bottom Row)} Tangential velocity of the preimage, which is universally
supersonic at the beginning, and becomes transsonic by \tsing.}
\label{fig.velhair} \end{center} \end{figure}

%% file: fig_r1000.tex
\begin{figure} \begin{center}
\includegraphics[width=\hw\textwidth]{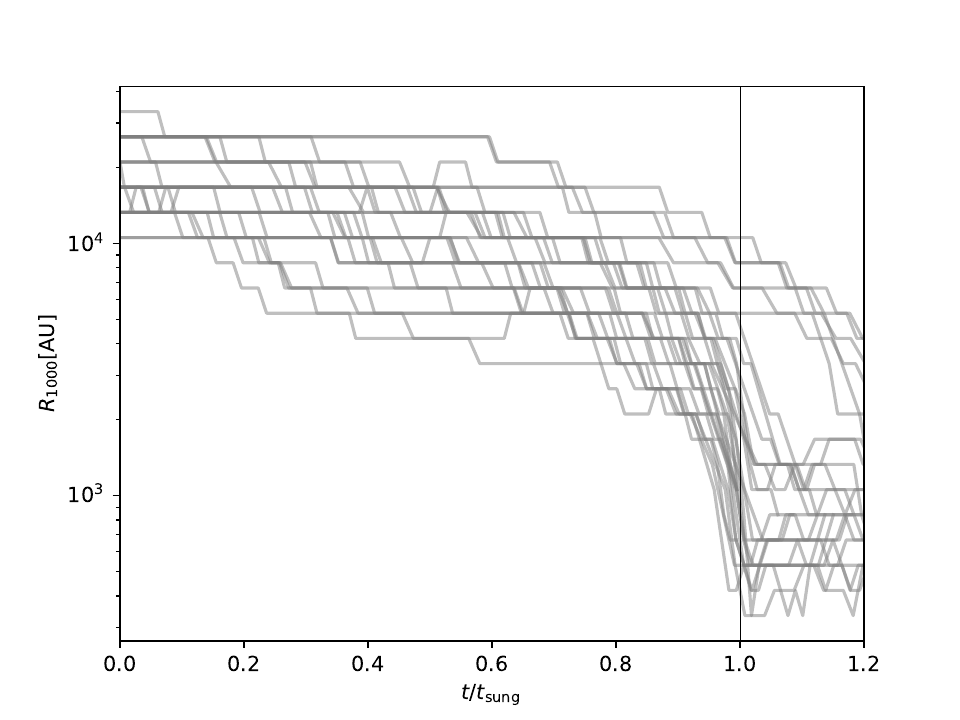}
\caption[ ]{$R_{1000}$ is the radius containing a mass of $M_{1000}$, which is
defined as the total mass within 1000 AU at the end of the simulation. 
  Most cores deliver the bulk of their mass to the center within 0.1 \tff\
of \tsing (see Figure \ref{fig.tsing}).}  
\label{fig.r1000} \end{center} \end{figure}

%% file: fig_freefall.tex
\begin{figure} \begin{center}
\includegraphics[width=\hw\textwidth]{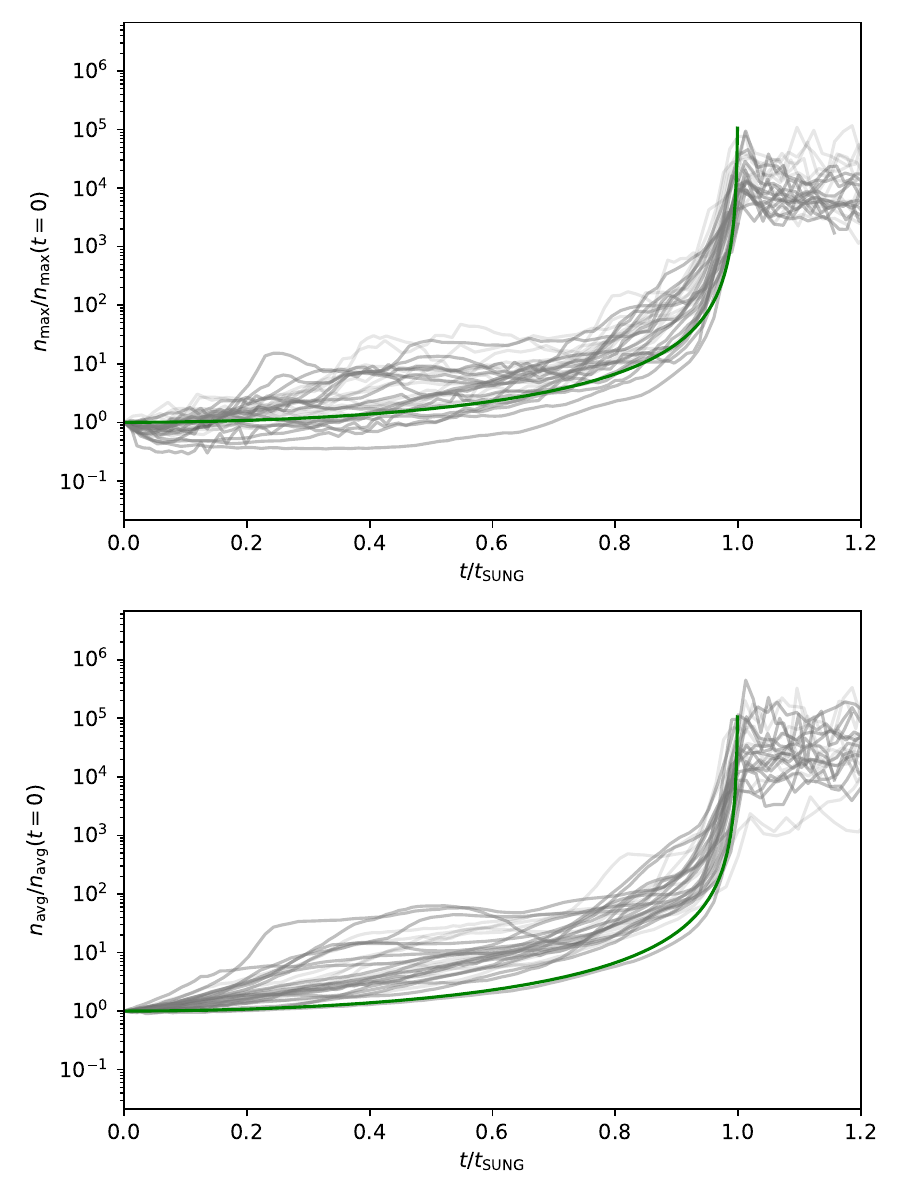}
\caption{(\emph{Top}) Peak density for each core vs. time normalized to the
initial maximum density and stretched in time to \tsung.  The green line shows
the free fall solution, Equation \ref{eq.freefall}.
 (\emph{Bottom}) Average density for each core, scaled to the initial average
 value and stretched in time.}
\label{fig.freefall} \end{center} \end{figure}

%% file: fig_fractal_time.tex
\begin{figure} \begin{center}
\includegraphics[width=\hw\textwidth]{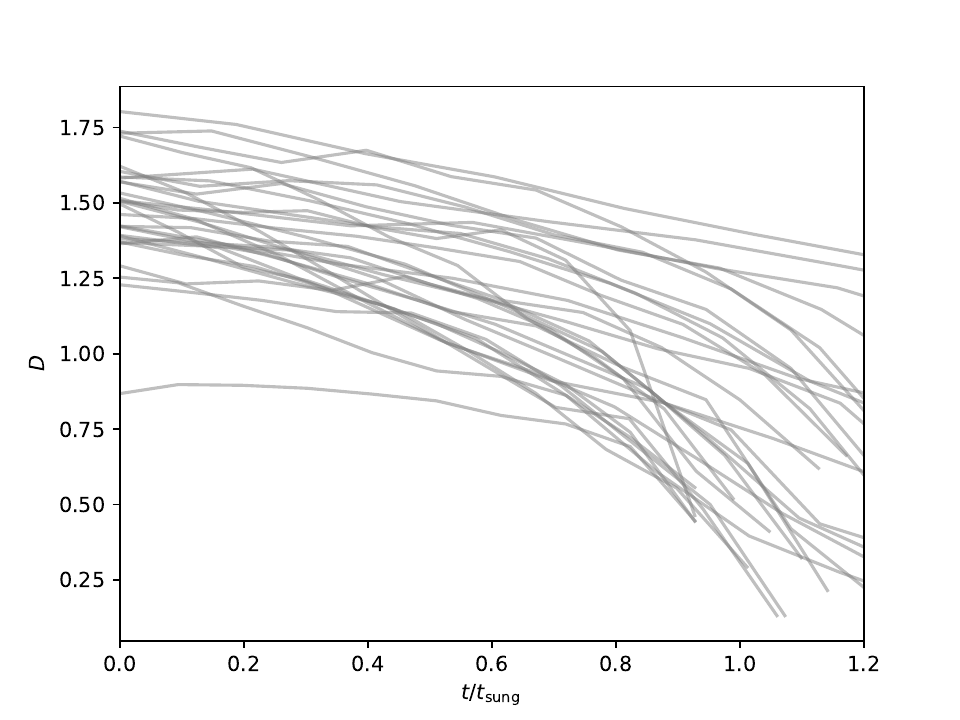}
\caption{Fractal dimension, $D$, vs time normalized to the end of the collapse,
\tsung. }
\label{fig.fractal} \end{center} \end{figure}

%% file: fractal_time.tex
\subsubsection{Fractal dimension}
\label{sec.fractal}

In Paper I, we examined the fractal dimension of the preimage gas at the initial
snapshot.  We use the Minkowski (box counting) dimension, defined as
\begin{align}
    D = \lim_{\epsilon \to 0} \frac{\log N(\epsilon)}{\log 1/\epsilon}.
\end{align}
$N(\epsilon)$ is the number of boxes of size $\epsilon$ that are needed to cover
the preimage gas as $\epsilon$ decreases.
We find in the first paper that the distribution of dimensions for the preimage
gas spreads between $D=0.25$ and $D=2$, peaked at 1.6.

Figure \ref{fig.fractal} shows $D$ vs. time for each of the single cores from
\sima, normalized to the collapse time \tsung.  Here, a grid at the
coarsest resolution ($\epsilon=1/128$ in code units) was used to cover each
grid, and $D$ is computed by increasing $\epsilon$.  As the collapse proceeds,
$D$ decreases nearly monotonically for every core.  This is not surprising, as
gravity tends to reduce the dimensionality of a collapsing object: a triaxial
object will first
collapse along the shortest axis forming a pancake, and then along the next
shortest axis to form a filament, and then along the filament to form a knot
\citep{Zeldovich70}.  This is mirrored in the fractal dimension behavior, as the
dimension decreases.

%% file: fig_radial_prof.tex
\begin{figure*} \begin{center}
	\includegraphics[width=\textwidth]{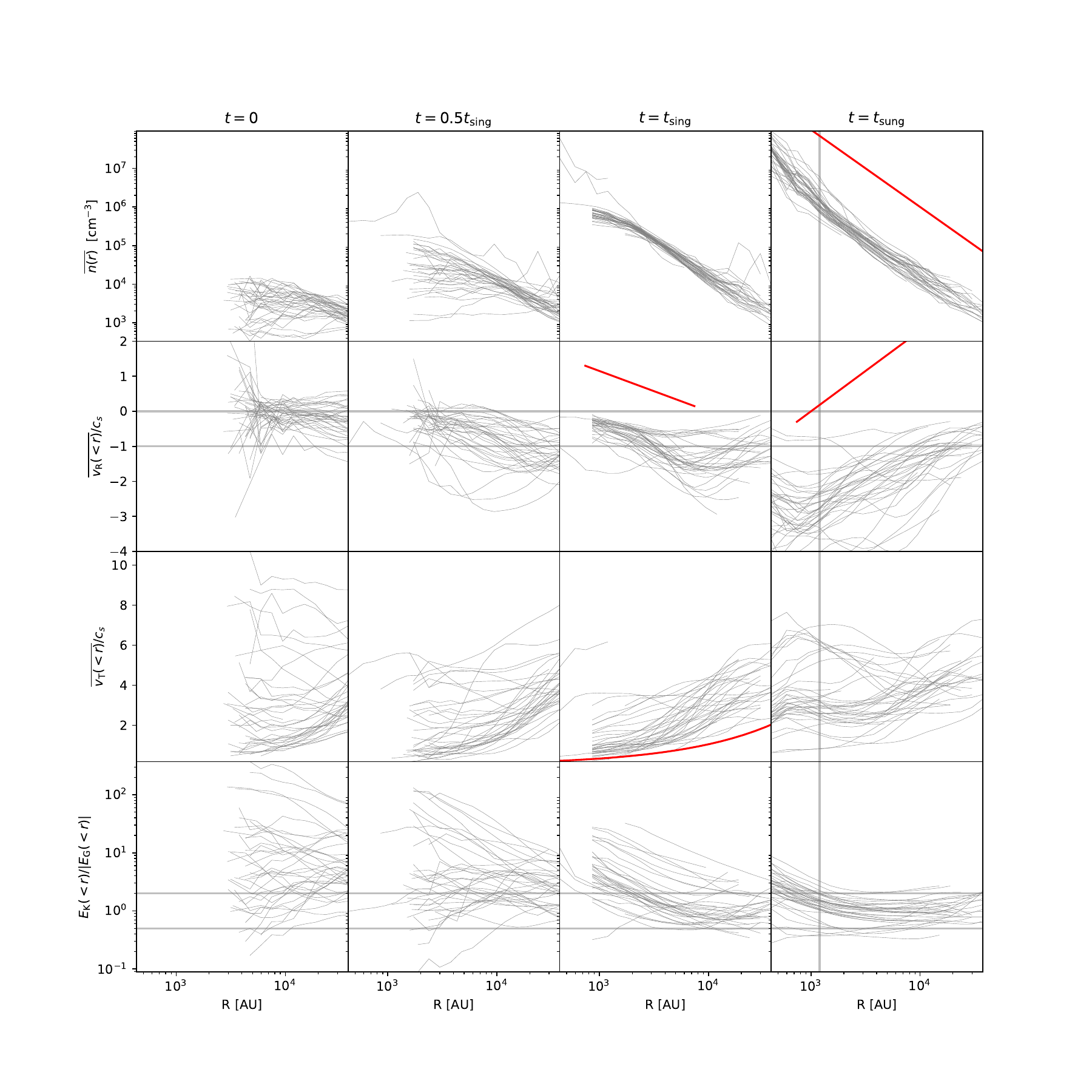}
\caption[ ]{Radial profiles for several quantities (rows) at several frames
(columns) for each single core (grey lines).  Four times at t=0, $t=0.5\tsing$, \tsing, and
\tsung are shown.  Note that \tsing\ and \tsung are different for each core.
The rows show mean density, radial velocity, tangential velocity, and ratio of
gravitational to kinetic energy.  Notable
features are highlighted by red lines, see text for details. }
\label{fig.radials} \end{center} \end{figure*}

%% file: radials.tex
\subsubsection{Radial Profiles at \tsing\ and \tsung}
\label{sec.radials}
\def\vrbar{\ensuremath{\overline{v_r}(r)}}
\def\vtbar{\ensuremath{\overline{v_t}(r)}}

Figure \ref{fig.radials} shows azimuthally  averaged quantities for spheres around every core (grey lines) at
four different times (columns).  The four columns show $t=0$, 
$0.5 \tsing$, $\tsing$, and $\tsung$.  Note that $\tsing$ and $\tsung$ are
different for every core, see Section \ref{sec.tsing} for definitions and
distributions.  The rows show azimuthal averages vs. radius for density, radial
velocity, tangential velocity, and ratio of energies.  Horizontal lines mark
relevant quantities; radial velocities of 0 and -1 and energy ratios of 0.5 and
2.  The vertical line in the last column shows the approzimate radius of the
pseudo-core, inside of which the result is dominated by numerics.

The first row shows azimuthally averaged density, 
\begin{align}
	\overline{n}(r) = \frac{1}{dV(r)} \int_{dV(r)} n(\xvec) d^2 \Omega
\end{align}
for every core at four
frames.  Here $V(r)$ is a sphere centered on the centroid of the particles with
radius $r$, and $dV(r)$ is its surface.  At $t=0$, density is roughly flat, showing a mild tendency for increased
density at the center. As discussed in \citet{Collins23},
this is a result of the density correlation length being significantly shorter
than the size of the preimage gas.  By $0.5\tsing$, a centrally condensed object
is beginning to form, some of them have developed an outer region that follows a
$\overline{n}\sim r^{-2}$
behavior.  By \tsing, the central densities have grown somewhat and all objects
begin to resemble Bonner-Ebert spheres \citep{Ebert55,Bonnor56}.
In the small
window between \tsing\ and \tsung, free-fall collapse sets in and the remaining
envelope is roughly $\overline{n}\sim r^{-2}$ as predicted by \citet{Penston69}.
The red line in the last panel shows $r^{-2}$ for comparison.

The second row of Figure \ref{fig.radials} shows mean interior radial velocity,
\begin{align}
    \overline{v_r}(r)=\frac{1}{V(r)}\int_{V(r)}
    (\vvec(\xvec)-\vvec_c)\cdot(\hat{r}_c) d^3\xvec
\end{align}
This is the average radial motion within a radius $r$, relative to the
central velocity for the core, $\vvec_c$, along the radial unit vector for that
core, $\hat{r}_c$.
At $t=0$, the radial velocity does not have a typical trend, some contracting
and some weakly expanding.  By the end of the \emph{collection} phase at $0.5
\tsing$, the radial velocity is negative for all particles.  Note that at early
phases, the curve of \vrbar\ does not, in general, go to zero as
$r\rightarrow 0$. This is due to the fractal and chaotic nature of the preimage
gas; the geometric center, the center of mass, and the center of velocity are
not at the same point in space.  By \tsing, 
the radial velocity tends to be logarithmic in radius, and subsonic.  We
posit that this phase is what is observed as a \emph{coherent core}.  Between \tsing\ and
\tsung, the majority of the mass is delivered to the center by negative radial
velocity.  After the relaxation
event, at \tsung, the radial velocity behavior is now increasing with radius,
and generally supersonic:
\begin{align}
\vrbar \propto 
\begin{cases}
	\ln \left(\frac{r}{r_0}\right)^{-1/2} & t=\tsing \label{eqn.vslope1}\\
	\ln \left(\frac{r}{r_0}\right)^{+1} & t=\tsung .
\end{cases}
\end{align}
This reversal happens by way of a wave that travels inwards.  The wave follows
the knee in the density profile. The red lines in the third and fourth panel
show the approximate scaling indicated by Equation \ref{eqn.vslope1}.

The third row of Figure \ref{fig.radials} shows tangential velocity,
\begin{align}
    \vtbar = \frac{1}{V(r)} \int_{V(r)} ||( (\vvec(\xvec)-\vvec_c)-\vvec_r)|| d^3 \xvec
\end{align}
which is just the full velocity less the radial in the frame of the core.
Also plotted in the third panel is a typical turbulent velocity expected from
theory; for an r.m.s. Mach number of 9, 
$v_{\rm{turb}}=9(r/L_0)^{1/2}$ \citep{Kritsuk07, Collins23}.  This can be seen
as the red line in the third panel.  Core profiles are bounded below by
this turbulent velocity.  During the hardening, the tangential velocity
maintains roughly a profile like $r^{1/2}$, decreasing in magnitude.  
After the
singularity, the rotation of the pseudo-core   
becomes quite large and no longer centered on zero velocity.  

In the fourth row of Figure \ref{fig.radials} we show the ratio of kinetic to gravitational
energy.  
We define gravitational energy $\EG$ and kinetic energy
$\EK$ as in Equation \ref{eqn.EGEK}
and we plot the ratio of their magnitudes. In the first panel, at $t=0$, we find that the ratio
tends to decrease with radius (though not universally).  This can be understood again as a signature of the
turbulent initial conditions.  The kinetic energy, $\EK$, increases as roughly
$r$, since the total velocity is subject to turbulent statistics, and
$v(r)\propto r^{1/2}$.  On the other hand, we integrate over several density correlation
lengths, so that average density is roughly constant and tends to the mean density.
Thus their ratio is roughly
linear in radius.  
This is a little counter intuitive, and can be understood because one preimage
region covers many density fluctuations, so there is on average no net
acceleration vector. During the hardening phase, $\EG$ and $\EK$ both increase
until they are more-or-less equal.  Large values of $\EK/\EG$ are seen at the
center, while the ratio adjusts to be between 0.5 and 2 by 10,000 AU.  During the singularity
between \tsing\ and  \tsung, the material in the pseudo-core begins to rotate,
and the ratio of energies in the envelope aligns to be roughly 2 at the center, between 0.5 and 2 for larger
radii.

%% file: conclusions.tex
\section{Conclusions}
\label{sec.conclusions}

%\red{Major Results:}
%\begin{itemize}
%\item Stars form faster than free-fall.  The exact source of the rate is not yet
%obvious, as it is broadly distributed and not correlated with the obvious
%contenders for rate.
%\item Stars do not begin as subsonic patches in the origin cloud.  Subsonic
%cores are seen, but they develop over time as the initially supersonic velocity
%decays.
%\item The pseudo-star assembles its mass quite quickly, and the remaining region
%around the pseudo-star
%\end{itemize}

In this work, we examine how the high vacuum of outer space becomes stars.  We
did this by embedding a fleet of pseudo-Lagrangian tracer particles in a
turbulent, collapsing molecular cloud; identifying dense cores and the particles
within; and then examining the density, velocity, and energetics of each
particle as they collapse.  

We find that collapse happens in three stages; \emph{collection},
\emph{hardening}, and \emph{singularity}.  The specifics are unique
to each core, and many cores do not experience the collection stage.  During
\emph{collection}, low density, high velocity gas is swept into moderate density
object by converging flows.  Sometimes these are initially dominated by gravity, sometimes by
kinetic energy.  During the \emph{hardening} phase, gravitational and kinetic
energy both increase to rough equipartition and the velocity decays to near
transsonic values.  During the \emph{singularity}, the
density gets very high, mass is delivered to the central object.  This is a universally fast event, lasting $\sim 0.1
\tff$.  At the end of singularity, a relaxed envelope remains, with
kinetic-to-gravitational energy ratio that varies logarithmically from about 2 at the edge of the pseudo-core
to about 0.5-2 at 10,000 AU.  We demonstrate this in detail for one representative
object in Section \ref{sec.case_study}, and the variation in mean quantities in
Section \ref{sec.ensemble}.

This can be seen as two related transitions to coherence: one in
which velocity of the gas that will collapse becomes low; and one in which the
kinetic and gravitational energies equilibrate.  The first is analogous to those
subsonic or trans-sonic cores  \citep{Pineda10, Singh21, Choudhury21}.  The
second is the fast equillibration of energies in the envelope.

Our study shows similarities with other recent works.  The three phase view of
\citet{Offner22} used machine learning techniques on similar simulations to
organize the flow.  They find that core experience three phases, turbulence,
coherence, and collapse, which are similar in spirit to our collection,
hardening, and singularity.  In their work, cores move between all phases, while our
cores do not.  One particular difference is our selection bias. We only examine
cores that do not interact with other dense objects.   Similarly the inertial flow model
\citep{Pelkonen21} has high velocity
streamers delivering the bulk of the mass to the core, which can be seen during
the singularity stage in our work.  These works are quite similar in setup and
simulation software to our own, differences in conclusions are due mostly to
differences in the lenses we use to view our simulations.  
The physics of self-gravitating turbulence is central to astrophysics, but
underpinned by chaos, so many views are necessary to unpack the true dynamics.

This work is a low resolution pilot to gain basic insight into this technique
and what it might show us.  Simulations with higher resolution and improved sink
particles are on the way.  Here we have made an attempt to only present results
that will stand up to improved resolution.  By improving resolution, one
increases the power of both density and velocity at the small scales.  Small
scale density power is not likely to affect our results, as we have shown many
density fluctuations are gathered together in the collapse process.  The increased velocity power, on
the other hand, is known to delay star formation, so our discussion of the
distribution of \tsing\ will be affected.  It is unlikely that more
resolution will introduce any features into this essentially featureless
distribution. The most that increased resolution will do is delay the first
collapse, shifting the distribution to higher $\tsing.$  It is also unlikely to alter the duration of the collapse as this
process is dominated by gravity which only depends on the integrated mass.  More
resolution will certainly allow the collapse to reach higher densities, but
again the error in the collapse duration will be small, as the time derivative
of density is also singular.  The fractal nature of the preimage gas will also be
unchanged, as further resolving a fractal yields more fractal, not less, by the
nature of a fractal.

%% file: data.tex
\section*{Data Availability}
Simulation data presented here is available on request (dccollins@fsu.edu).

%% file: acknowledgements.tex
\section*{Acknowledgements}
% Entry for the table of contents, for this guide only
\addcontentsline{toc}{section}{Acknowledgements}

We thank the referee for his careful feedback that greatly improved this work.
Support for this work was provided in part by the National Science Foundation
under Grant AST-1616026.  Simulations were performed on \emph{Stampede2}, part
of the Extreme Science and Engineering Discovery Environment
\citep[XSEDE;][]{Towns14}, which is supported by National Science Foundation grant number
ACI-1548562, under XSEDE allocation TG-AST140008.